\documentclass[twocolumn,pre,floatfix,superscriptaddress,nofootinbib]{revtex4-2}

\usepackage[T1]{fontenc}
\usepackage[latin9]{inputenc}
\usepackage{mathrsfs}
\usepackage{amsmath}
\usepackage{amssymb}
\usepackage{graphicx}
\usepackage{braket}
\usepackage[svgnames]{xcolor}   
\usepackage{comment}
\usepackage{natbib}
\usepackage{pifont}
\usepackage{algorithmic}
\usepackage{todonotes}

\usepackage{placeins}           

\makeatletter

\newcounter{algorithm}
\renewcommand{\thealgorithm}{\arabic{algorithm}}

\@namedef{theHalgorithm}{\thealgorithm}
\newenvironment{algorithm}[1][]{%
  \par\addvspace{\medskipamount}%
  \noindent\begin{minipage}{\linewidth}%
    \hrule height 0.6pt\vspace{2pt}%
    \renewcommand{\caption}[1]{%
      \refstepcounter{algorithm}%
      \vspace{2pt}\hrule height 0.4pt\vspace{2pt}%
      \noindent\textbf{Algorithm~\thealgorithm:} ##1\par}%
}{%
  \end{minipage}\par\addvspace{\medskipamount}%
}
\makeatother                    

\usepackage{listings}
\usepackage{hyperref}
\hypersetup{
  colorlinks=true,
  linkcolor=blue,
  citecolor=ForestGreen,
  urlcolor=DarkOrchid
}
\usepackage{cleveref}   

\newcommand{\HP}{\mathbb{H}}

\newcommand{\etal}{\textit{et al.}}
\newcommand{\qReduMIS}{\textsf{qReduMIS}}
\newcommand{\graph}{\mathcal{G}}
\newcommand{\Qset}{\mathcal{Q}}

\usepackage{amsthm}

\begin{document}

 \title{qReduMIS: A Quantum-Informed Reduction Algorithm for the Maximum Independent Set Problem}

\author{Martin J. A. Schuetz}
\thanks{These authors contributed equally.}
\affiliation{Amazon Advanced Solutions Lab, Seattle, Washington 98170, USA}
\affiliation{AWS Center for Quantum Computing, Pasadena, CA 91125, USA}

\author{Romina Yalovetzky}
\thanks{These authors contributed equally.}
\affiliation{Global Technology Applied Research, JPMorgan Chase, New York, NY 10017 USA}

\author{Ruben S. Andrist}
\affiliation{Amazon Advanced Solutions Lab, Seattle, Washington 98170, USA}

\author{Grant Salton}
\affiliation{Amazon Advanced Solutions Lab, Seattle, Washington 98170, USA}
\affiliation{AWS Center for Quantum Computing, Pasadena, CA 91125, USA}

\author{Yue Sun}
\affiliation{Global Technology Applied Research, JPMorgan Chase, New York, NY 10017 USA}

\author{Rudy Raymond}
\affiliation{Global Technology Applied Research, JPMorgan Chase, New York, NY 10017 USA}

\author{Shouvanik Chakrabarti}
\affiliation{Global Technology Applied Research, JPMorgan Chase, New York, NY 10017 USA}

\author{Atithi Acharya}
\affiliation{Global Technology Applied Research, JPMorgan Chase, New York, NY 10017 USA}

\author{Ruslan Shaydulin}
\affiliation{Global Technology Applied Research, JPMorgan Chase, New York, NY 10017 USA}

\author{Marco Pistoia}
\altaffiliation[]{These authors acted as Co-PIs.}
\affiliation{Global Technology Applied Research, JPMorgan Chase, New York, NY 10017 USA}

\author{Helmut G. Katzgraber}
\altaffiliation[]{These authors acted as Co-PIs.}
\affiliation{Amazon Advanced Solutions Lab, Seattle, Washington 98170, USA}

\date{\today}

\begin{abstract}

We propose and implement a quantum-informed reduction algorithm for the maximum independent set problem that integrates classical kernelization techniques with information extracted from quantum devices.
Our larger framework consists of dedicated application, algorithm, and hardware layers, and easily generalizes to the maximum weight independent set problem. 
In this hybrid quantum-classical framework, which we call qReduMIS, the quantum computer is used as a co-processor to inform classical reduction logic about frozen vertices that are likely (or unlikely) to be in large independent sets, thereby opening up the reduction space after removal of targeted subgraphs.   
We systematically assess the performance of qReduMIS based on experiments with up to 231 qubits run on Rydberg quantum hardware available through Amazon Braket.
Our experiments show that qReduMIS can help address fundamental performance limitations faced by a broad set of (quantum) solvers including Rydberg quantum devices. 
We outline implementations of qReduMIS with alternative platforms, such as superconducting qubits or trapped ions, and we discuss potential future extensions. 

\end{abstract}

\maketitle

\section{Introduction}
\label{sec:introduction}

The maximum independent set (MIS) problem is a paradigmatic, NP-hard combinatorial optimization problem with close ties to the well-known, complementary maximum clique and minimum vertex cover problems \citep{garey:90}. 
Practical applications can be found in virtually every industry, including computer vision \citep{feo:94}, map labeling \citep{gemsa:16}, network design \citep{hale:80}, vehicle routing \citep{dong:22}, and quantitative finance \citep{boginski:05, kalra:08, hidaka:23}, to name a few.
Given an undirected graph $\graph = (\mathcal{V}, \mathcal{E})$ with nodes $\mathcal{V}$ and edges $\mathcal{E}$, an independent set is a subset $\mathcal{I} \subseteq \mathcal{V}$, such that no two vertices in the set $\mathcal{I}$ share an edge. 
The goal of the MIS problem is to find an independent set $\mathcal{I}$ with maximum cardinality.  
Such a set is called a \textit{maximum independent set} of size $|\mathrm{MIS}|=|\mathcal{I}|$.
Similarly, for a weighted graph $\graph = (\mathcal{V}, \mathcal{E}, \omega)$ with vertex weight function $\omega:\mathcal{V} \rightarrow \mathbb{R}^{+}$, the generalized maximum weight independent set (MWIS) problem asks for an independent set $\mathcal{I}$ with maximum weight $\omega({\mathcal{I}})=\sum_{v \in \mathcal{I}}\omega(v)$.

\textbf{Classical ReduMIS.} One of today's leading heuristics for the MIS problem is the ReduMIS algorithm \citep{lamm:17}, which intermixes a suite of reduction (or kernelization) techniques with a heuristic evolutionary algorithm. 
Reduction techniques are polynomial time procedures that can shrink a given input graph to an irreducible kernel by removing well-defined subgraphs. These subgraphs are removed through targeted selection of exposed vertices that are provably part of some maximum(-weight) independent set, hence maintaining optimality \citep{butenko:02, butenko:07, butenko:09, strash:16, chang:17, hespe:19, grossmann:24}.  
Whenever an irreducible kernel is identified, ReduMIS makes use of an evolutionary approach to identify and remove vertices that are likely part of a large independent set, thereby opening up yet again the reduction space for further kernelization. Ultimately, a solution to the MIS or MWIS problem on the original input graph can be found by undoing previously applied reductions, with numerical experiments reporting state-of-the-art results for both large MIS and MWIS problem instances \citep{lamm:17, grossmann:23}.

\textbf{Quantum optimization.} Over the last few decades, quantum optimization algorithms have emerged as a novel paradigm for solving combinatorial optimization problems, such as the MIS problem \citep{abbas:23}. 
Prominent examples include quantum annealing algorithms (QAA) \citep{kadowaki:98, farhi:00, farhi:01, das:08, hauke:20} and the quantum approximate optimization algorithm (QAOA) \citep{farhi:14, zhou:20}.
In particular, analog neutral-atom quantum machines in the form of Rydberg atom arrays have attracted broad interest as a novel class of programmable and scalable special-purpose quantum devices that can natively encode and (approximately) solve the MIS (and MWIS) problem on unit-disk (UD) graphs \citep{pichler:18, pichler:18computational, serret:20, ebadi:22, cain:23, schiffer:23, finzgar:23, finzgar:23b, perseguers:24}.
Recent QAA-based experiments report on a potential super-linear quantum speedup over classical simulated annealing \citep{ebadi:22, andrist:23}. 
However, for a broad set of Markov-chain Monte Carlo (MCMC) algorithms, the same experiments also show that the algorithmic performance is suppressed exponentially in the conductance-like hardness parameter $\HP$, defined as $\HP =D_{\mathrm{|MIS|-1}}/(|\mathrm{MIS}|\cdot D_{|\mathrm{MIS}|})$, where $D_{\alpha}$ denotes the degeneracy of the independent sets of size $\alpha$.
Specifically, the success probability to find the MIS in a single algorithmic run (shot), denoted as $P_{\mathrm{MIS}}$, is shown to follow the form $P_{\mathrm{MIS}} \approx 1 - \exp(-C \HP^{-\beta})$ \citep{ebadi:22}, where $\beta$ depends on details of the given algorithm, and $C$ refers to a positive (fitted) constant.

\begin{figure*}
  \includegraphics[width=1.85 \columnwidth]{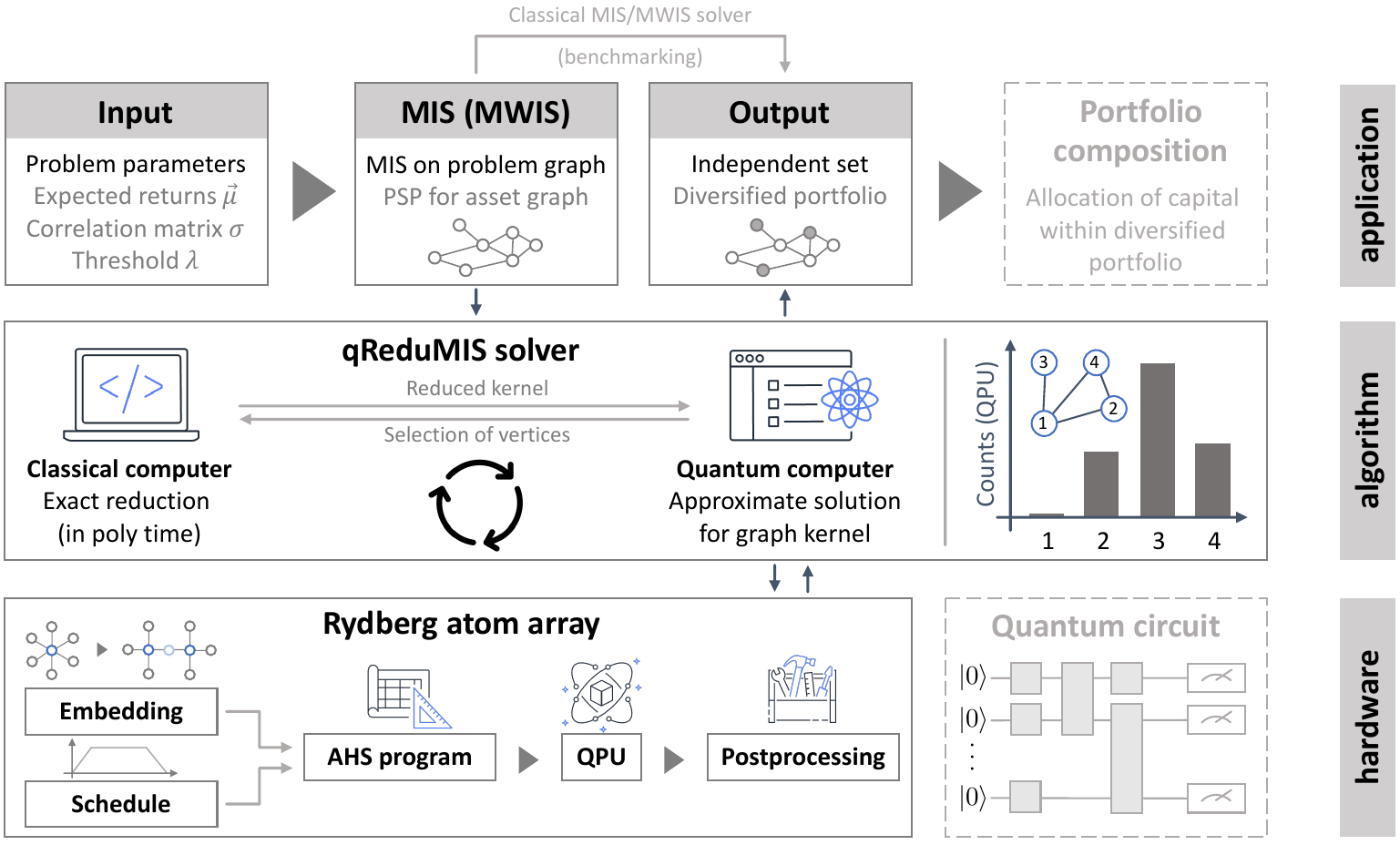}
  \caption{Schematic illustration of the hybrid {\qReduMIS} framework, with distinct (i) application, (ii) algorithm, and (iii) hardware layers (from top to bottom). 
  The application layer is illustrated for an example portfolio selection problem (PSP) that can be framed as a MIS (or MWIS) problem on an asset graph, the solution to which can be fed into a downstream portfolio composition problem. 
  For a given problem instance on a graph $\mathcal{G}$, the {\qReduMIS} algorithm solves the MIS problem by intermixing exact, polynomial time reduction logic with information obtained from quantum measurements to unblock the kernelization process whenever necessary by identifying and removing frozen vertices that have a high (or low) likelihood to be part of large independent sets. 
  The {\qReduMIS} framework is hardware-agnostic, and can be powered by both \textit{digital} or \textit{analog} quantum devices, such as Rydberg atom arrays implementing analog Hamiltonian simulation (AHS) programs.      
      \label{fig:scheme}
  }
\end{figure*}

\textbf{Quantum ReduMIS (qReduMIS).} In this work we propose and implement a novel quantum-informed algorithm (dubbed \textit{qReduMIS}), which integrates exact classical kernelization with a quantum processing unit (QPU) used as co-processor (rather than a classical evolutionary heuristic) that unblocks the reduction process whenever necessary;  
see Fig.~\ref{fig:scheme} for a schematic illustration.
Unlike standard quantum optimization
approaches---which task the QPU with sampling a maximum independent
set directly---qReduMIS uses the QPU solely to identify
\emph{frozen nodes}: vertices whose marginal probability of belonging
to large independent sets is anomalously high (or low).  These frozen
nodes are then removed (along with their neighbourhoods),
\emph{re-enabling} exact classical reduction rules on the resulting
kernel.  This process is \emph{recursive}: the algorithm alternates
between classical kernelization and QPU-assisted frozen-node
identification until the graph is fully reduced or a target kernel
size is reached.
The algorithm is hardware-agnostic and embedded in a larger,
three-tiered framework, with distinct application, algorithm, and
hardware layers.
Because each QPU call operates on a progressively
smaller kernel, qReduMIS provides a path towards smaller and simpler
kernels that are more accessible to the QPU. While this need not hold
for very large instances, the sampling quality of the quantum device
generically remains high, even when the original instance may be far
too hard for standard quantum optimization to sample an optimal
solution.
By design, the algorithm leans heavily on classical kernelization, 
and periodically leverages information obtained from QPU measurements to unblock the kernelization process by identifying and removing frozen vertices that have a high (or low) likelihood to be part of large independent sets.
This offers several benefits: First, classical kernelization is
provably optimal, second these methods are fast (e.g., linear-time
for sparse graphs), and third, kernelization methods are insensitive
to the hardness parameter $\HP$ \citep{schuetz:24}.

We emphasize that the classical reduction logic in qReduMIS is
\emph{not} merely a one-time preprocessing step applied before handing a
(smaller) problem to a quantum solver.  This distinguishes qReduMIS
from warm-start approaches~\cite{egger2021warm, tate2023bridging}, which
classically pre-compute an initial state or parameter set and then
run a \emph{full-size} quantum optimization (e.g., QAOA) as a second,
independent stage.  In qReduMIS, reduction and quantum sampling are
interleaved in a recursive loop: the QPU output feeds back into the
reducer, which produces a new (smaller) kernel, which may require
another QPU call.  Classical reduction is therefore an integral,
repeated component of the algorithm---not an offline preprocessing
step---and generically the QPU never sees the full original graph.

Our contributions can be summarized as follows:
We introduce reduction (kernelization) techniques within
a recursive (\textit{hybrid}) framework, thereby
effectively expanding the size of problems compatible with near-term
quantum hardware within larger end-to-end pipelines.
Disentangling the reduction logic from the kernel solver (which can be classical or quantum) our framework is designed to be hardware-agnostic, thus unlocking new experiments across hardware platforms (including Rydberg atoms, trapped ions, and superconducting qubits). Our experiments with up to 231 qubits demonstrate that qReduMIS successfully tackles fundamental performance issues reported in Refs.~\cite{ebadi:22, andrist:23, perseguers:24}, significantly outperforming previous approaches based on optimized quantum annealing protocols \cite{ebadi:22, perseguers:24}.

\section{$\text{qReduMIS}$ Methodology}
\label{sec:methodology}

In the following, we outline the anatomy of qReduMIS in more detail, systematically stepping through its individual layers from application, to algorithm, to hardware layers (cf.~Fig.~\ref{fig:scheme} from top to bottom). 

\textbf{Application layer.} The qReduMIS algorithm accepts generic MIS instances as problem input and can be extended to MWIS instances by combining generalized reduction techniques for MWIS \citep{grossmann:23} with quantum devices compatible with the MWIS problem \citep{nguyen:23, deoliveira:24}.
Generically, in the application layer we construct a MIS~/~MWIS problem instance for a given set of problem-specific input parameters. The corresponding problem graph is fed to the algorithm layer which then returns a (large) independent set that can be further processed based on the underlying use case. 
For illustration, in Fig.~\ref{fig:scheme} we showcase the application layer with an example portfolio selection problem (PSP). 
Given information from historical stock price data in the form of expected returns $\boldsymbol{\mu}$ and inter-asset correlations $\sigma$, together with a user-defined correlation threshold $\lambda$, the PSP can be framed as an MIS (or MWIS) problem on a so-called asset (or market) graph \citep{kalra:08, hidaka:23, macMahon:15}, 
which is passed to the algorithm layer. 
The output solution represents a diversified portfolio within a larger universe of assets which can be integrated into a larger multi-stage portfolio management pipeline 
where portfolio selection is followed by portfolio composition \citep{hidaka:23}.  

        \begin{algorithm}[t]
                \begin{algorithmic}
                \small
                        \STATE   \textbf{Input} graph $G=(V,E)$, RCL size $K_{\mathrm{RCL}}$, selection parameter $\lambda$ \textcolor{gray}{\# default $K_{\mathrm{RCL}}=\lambda=1$}
                        \STATE   \textbf{Global} $\mathcal{W}= \emptyset$, $\mathcal{S}= \emptyset$, $\mathcal{R}= \emptyset$ \textcolor{gray}{\# best solution (incumbent), set of selected nodes, set of removed nodes}
                        \STATE   \textcolor{gray}{\# Set up selection scheme}
                        \STATE   $\mathrm{params} = \{\mathrm{size}: \lambda, \mathrm{RCLsize}: K_{\mathrm{RCL}}, \mathrm{strategy}: in\}$ 	
                        \STATE   \textbf{Procedure} {\qReduMIS}($G$)
                        \STATE   \quad \textbf{if} $G$ is empty \textbf{then} \textbf{return} $\mathcal{W}$
                        \STATE   \quad \textcolor{gray}{\# Get kernel, selected, and removed nodes}
                        \STATE   \quad $(\mathcal{K}, s, r) \leftarrow$ \textsc{ClassicalReduce}$(G)$
                        \STATE   \quad Append $s$ to $\mathcal{S}$, append $r$ to $\mathcal{R}$
                        \STATE   \quad \textbf{if} $\mathcal{K}$ is empty \textbf{then} update and \textbf{return} $\mathcal{W}$
                        \STATE   \quad \textcolor{gray}{\# Get candidate sets for kernel from QPU}
                        \STATE   \quad $\{\mathcal{I}_{n}\} \leftarrow$ \textsc{QuantumMIS}$(\mathcal{K})$
                        \STATE   \quad \textcolor{gray}{\# Update incumbent}
                        \STATE   \quad \textbf{if} $|\mathcal{S}|+\mathrm{max}(|\mathcal{I}_{n}|)>|\mathcal{W}|$ \textbf{then} update $\mathcal{W}$
                        \STATE   \quad \textcolor{gray}{\# Select $\lambda$ frozen (fixed) nodes with "in/out" strategy,}
                        \STATE   \quad \textcolor{gray}{\# and update $\mathcal{S}$ and $ \mathcal{R}$ accordingly}
                        \STATE   \quad $(\Qset, s_{q}, r_{q}) \leftarrow \textsc{Select}(\{\mathcal{I}_{n}\}, \mathrm{params})$
                        \STATE   \quad Append $s_{q}$ to $\mathcal{S}$, append $r_{q}$ to $\mathcal{R}$
                        \STATE   \quad $\mathcal{K'}  \leftarrow \mathcal{K}[\mathcal{V}_{\mathcal{K}} \backslash \Qset]$ \textcolor{gray}{\# Get updated (inexact) kernel}
                        \STATE   \quad {\qReduMIS}($\mathcal{K'}$) \textcolor{gray}{\# Recurse on inexact kernel}
                        \STATE   \textbf{return} $\mathcal{W}$
                \end{algorithmic}
                \caption{Basic Structure of {\qReduMIS}}
                \label{algo:qReduMIS}
                \vspace*{-.1cm}
        \end{algorithm}

\textbf{Algorithm layer.} The algorithm layer is at the center of the qReduMIS framework, involving both classical and quantum resources, with the QPU feeding information to the classical reduction logic in every iteration of the algorithm. 
See Algorithm \ref{algo:qReduMIS} for a basic outline of the core routine. 
While several variants of qReduMIS are conceivable, ranging from greedy heuristics to exact branch-and-reduce-type algorithms, here we focus on a semi-greedy implementation thereof. 
Independent of implementation details, qReduMIS comprises three core modules, given by (i) $\textsc{ClassicalReduce}(\cdot)$, (ii) $\textsc{QuantumMIS}(\cdot)$, and (iii) $\textsc{Select}(\cdot)$ methods.

The $\textsc{ClassicalReduce}$ routine takes a graph $\mathcal{G}$ as input and outputs a smaller graph kernel $\mathcal{K}$, along with sets of nodes that have been selected or removed, respectively. 
To achieve this reduction, one can draw from the full suite of reduction procedures developed over the last two decades \citep{lamm:17, lamm:19, grossmann:24}, involving, for example, techniques such as isolated vertex removal \citep{butenko:02}, vertex folding \citep{chen:01}, or critical independent set reductions \citep{butenko:07}.
For definiteness, here we leverage a simple, yet efficient reduction procedure known as isolated vertex removal \citep{butenko:02, strash:16, schuetz:24}, which makes use of the notion of exposed corner nodes (also known as isolated or simplicial vertices) that are provably part of some MIS/MWIS by a simple cut-and-paste argument \citep{butenko:02, hespe:19, strash:16, schuetz:24}. 
By iteratively selecting those corner nodes and removing their neighborhood, this simple kernelization technique is able to efficiently shrink a given input graph $\mathcal{G}$ to an irreducible kernel $\mathcal{K}$---typically in linear time---while preserving optimality \citep{schuetz:24}.  
Here we will focus on the MIS problem, but it is straightforward to generalize qReduMIS to the MWIS problem, by either (heuristically) extending our reduction logic towards weighted nodes as detailed in Ref.~\cite{schuetz:24} or by adopting known (exact) MWIS reduction schemes \cite{grossmann:23}.

Next, the irreducible kernel $\mathcal{K}$ is passed to the $\textsc{QuantumMIS}$ routine, which samples from the low-energy sector and computes high-quality solutions in the form of independent sets $\{\mathcal{I}_{n}\}$ for the kernel, for $n=1, \dots, n_{\mathrm{shots}}$, with $n_{\mathrm{shots}}$ referring to the number of QPU measurements (shots). 
As further detailed below, both analog (i.e., quantum annealing type) or digital (i.e., circuit-based) quantum devices can be called in the backend to compute these independent sets. 
In the case that the kernel's size exceeds the QPU's capabilities, qubit compression \citep{sciorilli:25}, lightcone \citep{dupont:25}, or decomposition techniques \citep{acharya:24} may be used to effectively shrink the problem size and enable the use of near-term quantum devices.   

Finally, the $\textsc{Select}$ routine takes the QPU's output $\{\mathcal{I}_{n}\}$ as input, and leverages its inherent sampling capabilities to identify frozen nodes that tend to be included in independent sets (or excluded), across all measurements (or some suitably filtered version thereof). See the histogram in Fig.~\ref{fig:scheme} for a simple example with four nodes in which node $1$ is never part of a low-energy solution, while node $3$ is always selected (across different MIS solutions). In this case, both nodes $1$ and $3$ are \textit{frozen}. 
In the context of superconducting annealers, this concept has been introduced as \textit{sample persistence} \citep{karimi:17}, which amounts to the evaluation of one-point correlations sourced from a low-energy (quantum) state, but can be generalized to higher-order correlations \citep{finzgar:23b, bravyi:20}. 
Here, we consider two simple selection mechanisms, both based on a histogram of in-set selection counts across all nodes $i=1, \dots, |\mathcal{V}_{\mathcal{K}}|$, as can be straightforwardly sourced from the measurement results $\{\mathcal{I}_{n}\}$; compare Fig.~\ref{fig:scheme} for a toy illustration. 
We refer to \textit{in-set} (\textit{out-set}) nodes as those vertices which have the highest (lowest) count numbers, respectively, representing their likelihood to be part of high-quality independent set solutions.   
In a semi-greedy implementation of qReduMIS, we randomly pick $\lambda$ candidate nodes from a restricted candidate list (RCL) of size $K_{\mathrm{RCL}}$ that tracks those nodes with the highest (lowest) counts, possibly with a bias towards low-degree nodes for in-set nodes, simply because high-degree vertices are unlikely to be in a large independent set \citep{lamm:17}.
The nodes identified from the RCL are then used to construct a set $\Qset$ to remove from the kernel. For the out-set strategy,  $\Qset$ contains the $\lambda$ randomly-selected nodes, whereas for the in-set strategy $\Qset$ contains the $\lambda$ randomly-selected nodes \textit{and} their neighbors.
After this quantum-informed elimination of $\Qset$, we repeat the procedure on the updated kernel $\mathcal{K'}  \leftarrow \mathcal{K}[\mathcal{V}_{\mathcal{K}} \backslash \Qset]$, 
with the central idea that the quantum-informed removal of $\Qset$ once again opens up the reduction space for further classical kernelization and helps guide the algorithm towards an MIS solution.  

\textbf{Hardware layer.} The qReduMIS hardware layer can involve hardware solvers such as Ising machines \citep{mohseni:22}, and analog or digital quantum devices. 
Specifically, quantum algorithms such as QAA \citep{kadowaki:98, farhi:00, farhi:01, das:08, hauke:20} or QAOA \citep{farhi:14, zhou:20} run on quantum devices can be used to prepare a superposition of low-energy candidate solutions to the MIS/MWIS problem from which information is extracted upon measurement to unblock classical kernelization. 

Given a graph $\mathcal{K}=(\mathcal{V}_{\mathcal{K}}, \mathcal{E}_{\mathcal{K}})$, the MIS problem is equivalent to finding the ground state of the classical Hamiltonian 
\begin{equation}
H(\mathbf{n}) = -\sum_{i \in \mathcal{V}_{\mathcal{K}}}n_{i} + U\sum_{(i,j) \in \mathcal{E}_{\mathcal{K}}}n_{i}n_{j},
\label{eq:hamiltonian-classical}
\end{equation}
with $U>1$ and $n_{i}=1$ if node $i$ is in the set (and $n_{i}=0$ otherwise); generalization to the MWIS problem is straightforward. 
Introducing the Ising spin variables $z_{i}=2n_{i}-1$ and promoting those to Pauli spin operators $z_{i} \rightarrow \hat{Z}_{i}$ we obtain the corresponding MIS \textit{quantum} Ising Hamiltonian 
\begin{equation}
\hat{H}_{\mathrm{cost}}=\sum_{i,j}J_{ij}\hat{Z}_{i}\hat{Z}_{j}+\sum_{i}h_{i}\hat{Z}_{i},
\label{eq:hamiltonian-ising}
\end{equation}
with local fields $h_{i}$ and couplers $J_{ij}$.  
The common theme across different quantum algorithms and hardware platforms is then to prepare a quantum many-body system undergoing coherent, programmable dynamics, as described by the unitary evolution
\begin{equation}
|\Psi(\theta)\rangle = \mathcal{U}(\theta)|\Psi_{\mathrm{initial}}\rangle,
\label{eq:quantum-state}
\end{equation}
which seeks to prepare $|\Psi(\theta)\rangle$ via control parameters $\theta$ as to minimize the expectation value with respect to the cost Hamiltonian $\hat{H}_{\mathrm{cost}}$.
The unitary $\mathcal{U}(\theta)$ can be engineered with both analog and digital quantum devices, across hardware platforms such as superconducting qubits or trapped ions, as further detailed in Appendix~\ref{app:hardware}. 
In the following, we focus on implementations based on Rydberg atom arrays.  

\begin{figure*}
  \includegraphics[width=2.05 \columnwidth]{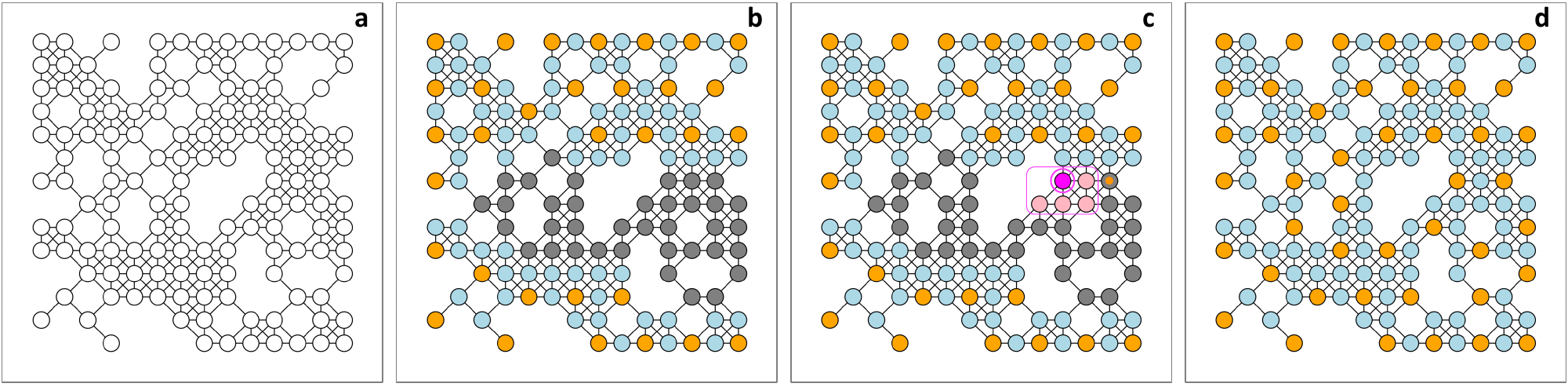}
  \caption{
  Snapshots of the qReduMIS algorithm (with in-set selection strategy) for the hardest instance listed in Tab.~\ref{tab:4hp} with $\HP \sim 1435$.
  \textbf{(a)} The problem input is given in terms of a site-diluted union-jack graph with $n=137$ nodes on a square lattice of side length $L=14$.  
  \textbf{(b)} qReduMIS first calls $\textsc{ClassicalReduce}$. Orange nodes are selected ($n_{i}=1$), blue nodes are removed ($n_{i}=0$) per kernelization, until an irreducible kernel $\mathcal{K}$ is found, with $37$ kernel nodes displayed in dark gray. 
  \textbf{(c)} Next, the QPU is called to unblock the reduction procedure. Solution candidates for the kernel are sampled, and the node highlighted in pink is identified as a frozen node with high in-set probability. 
  Selection of this node and removal of its kernel neighborhood (pink box) results in an updated kernel $\mathcal{K}'$ that features an exposed corner node (highlighted with an orange dot).   
  \textbf{(d)} The remaining kernel is fully reducible. 
  Within two iterations (with two calls to $\textsc{ClassicalReduce}$ and one quantum-informed selection) qReduMIS finds an optimal solution of size $|\mathrm{MIS}|=45$.  
  \label{fig:snapshots}
  }
\end{figure*}

Given the ability of Rydberg atom arrays to efficiently encode MIS (and MWIS) problems on unit-disk graphs \citep{pichler:18, ebadi:22}, here we focus on qReduMIS backed by Rydberg atom arrays.
See Fig.~\ref{fig:scheme} for a schematic illustration of our end-to-end workflow for solving the MIS with analog Hamiltonian simulation (AHS) programs run on Rydberg QPUs.  
In this setup, each atom represents a vertex, and coherent excitation from an atom's ground state $|0\rangle$ into an excited Rydberg state $|1\rangle$ can be utilized for hardware-efficient encoding of unit-disk MIS problems, with the Rydberg blockade mechanism effectively inducing edges within the (tunable) unit-disk radius $R_{b}\sim 1$ -- $10\mu\mathrm{m}$ \citep{adams:20, lukin:01, saffman:10, levine:19}.
The quantum dynamics of Rydberg atom arrays is described by the Hamiltonian $\hat{H}=\hat{H}_{\mathrm{mix}}+\hat{H}_{\mathrm{cost}}$, with ($\hbar=1$)
\begin{equation}
\label{eq:rydberg-hamiltonian}
\begin{aligned}
    \hat{H}_{\mathrm{mix}} &= \frac{\Omega(t)}{2} \sum_i \left(e^{i\phi(t)} |0\rangle_{i} \langle 1| + \mathrm{h.c.}\right),
    \\
    \hat{H}_{\mathrm{cost}} &=
    - \Delta(t) \sum_i \hat{n}_i
    +
    \sum_{i<j} V_{i j} \hat{n}_i \hat{n}_j,
\end{aligned}
\end{equation}
where $\hat{n}_i:=|1\rangle_{i} \langle 1|$ is the occupation number operator for atom $i$, $V_{ij}=C_6/\|\mathbf{x}_{i}-\mathbf{x}_{j}\|^{6}$ for atoms at positions $\mathbf{x}_{i}$ and $\mathbf{x}_{j}$, respectively, and $C_6$ is the van der Waals coefficient.  
Here, $\theta(t)=\{\Omega(t), \phi(t), \Delta(t)\}$ specifies a programmable schedule in terms of the global Rabi frequency $\Omega(t)$, the phase of the Rabi drive $\phi(t)$, and laser detuning $\Delta(t)$. 
Together with a two-dimensional atom arrangement $\{\mathbf{x}_{i}\}$, $\theta(t)$ defines a complete AHS program for quantum devices available online today with up to 256 qubits \cite{wurtz:23}.
Neglecting corrections from Rydberg interaction tails and carefully choosing positive detuning $\Delta_{\mathrm{UB}} > \Delta > \Delta_{\mathrm{LB}} >0$ \citep{perseguers:24}, it is straightforward to see that $\hat{H}_{\mathrm{cost}}$ in Eq.~(\ref{eq:rydberg-hamiltonian}) can capture the MIS cost function in Eq.~(\ref{eq:hamiltonian-classical}) on unit-disk graphs, in a quantum system with $n_{i} \rightarrow \hat{n}_{i}$, while the driver $\hat{H}_{\mathrm{mix}}$ allows to coherently drive transitions between bit strings \citep{pichler:18}. 
Generalized MWIS instances can be tackled using \textit{local} detuning capabilities, effectively taking $\Delta(t) \rightarrow \Delta_{i}(t)$ \citep{nguyen:23, deoliveira:24}.
If a given problem instance does not match the unit-disk connectivity supported natively by the hardware, we resort to embedding schemes, as described in Refs.~\citep{nguyen:23, schuetz:24}, that map the logical input to a physical graph compatible with the hardware connectivity constraints, typically at the expense of a quadratic overhead in the required number of qubits.
At the end of the schedule, site-resolved projective measurements can be used to read out the final quantum many-body state. 
Given this raw output, we use simple post-processing logic that repairs potential violations of the independence constraint and ensures the feasibility of all solution candidates \citep{ebadi:22}. 
Out-set (in-set) nodes to be considered for selection per $\textsc{Select}$ method then manifest as frozen (persistent) vertices for which the local observables $\hat{n}_{i}$ have low (high) expectation values, i.e., $\langle \hat{n}_{i} \rangle \approx 0$ ($\langle \hat{n}_{i} \rangle \approx 1$), 
as those give the probability to measure the $i$-th atom in the Rydberg state $|1\rangle$ at the end of the anneal.

\begin{table}
\caption{
Average success probabilities $P_{\mathrm{MIS}}$ achieved with QAA and qReduMIS (with in-set selection strategy) using the QuEra Aquila QPU for four test instances (all with $n=137$ vertices) referenced by their hardness parameters $\HP$.
More details can be found in Appendix~\ref{additional-numerics}.
\label{tab:4hp}} 
\begin{tabular*}{\columnwidth}{@{\extracolsep{\fill}} l c c c c }
\hline
\hline
 Instance hardness $\HP$ & $\sim 1.478$ & $\sim 14.12$ & $\sim 125.5$ & $\sim 1435$ \\ [0.5ex] 
\hline
 QAA & 35.2\% & 0.8\% & 0.7\% & 0\%   \\ 
 QAA (Ref.~\citep{perseguers:24}) & 47.8\% & 0.4\% & 1.9\% & 0\%  \\ 
 qReduMIS & 100\% & 100\% & 100\% & 100\%  \\ [1ex] 
\hline
\hline
\end{tabular*}
\end{table}

\textbf{qReduMIS run time.} 
Unless a maximum number of iterations is specified, qReduMIS  proceeds until the kernel is empty. We denote the number of iterations needed to fully reduce the kernel by $D$. Every iteration involves both classical ($\tau_\mathrm{cl}$) as well as quantum processing time ($\tau_\mathrm{q}$), up to the last step that only involves classical kernelization. 
Considering realistic wall-clock times, we have $\tau_\mathrm{cl} \ll \tau_\mathrm{q}$, leading to a total run time $T \approx D \cdot n_{\mathrm{shots}} \tau$, with $\tau_\mathrm{q} = n_{\mathrm{shots}} \tau$ and hardware-dependent cycle time $\tau$.
For current Rydberg devices, $\tau$ is limited by (destructive) blow-out measurement techniques to $\tau \sim 0.1\mathrm{s}$ \citep{wurtz:23}.
For typical input graphs with $n \sim 1000$ nodes, we estimate $D \lesssim 10$, based on repeated classical kernelization, leading to $T \approx 100\mathrm{s}$ -- $1000\mathrm{s}$ for $n_{\mathrm{shots}} \approx 100$ -- $1000$ measurements per iteration. 
This timescale can be reduced via faster cycle times through improved reloading schemes \citep{wurtz:23}, the ability to parallelize measurements across a single QPU device provided both enough space and qubits are available \citep{wurtz:23}, and the potential to parallelize the workload across multiple QPUs, if available. 

\textbf{Frozen-node sampling.} Rather than tasking the QPU with
sampling a maximum independent set, qReduMIS only uses it to identify
frozen nodes, i.e., nodes with a high (or low) probability of being in (or out of)
the set. This offers two key benefits.

\textit{Reduced problem size.} Classical reduction is exact and typically
shrinks the graph to a smaller kernel before any QPU call. Since
Ebadi \etal~\citep{ebadi:22} have shown that, for random unit-disk
instances, the hardness parameter $\HP$ grows on average with system
size, the successive (smaller) kernels handed to the QPU are
typically easier to sample than the original instance.

\textit{Robustness to errors.} By design, qReduMIS can find the MIS
of the original input $\mathcal{G}$ even when the QPU never samples an
MIS for the kernels $\mathcal{K}$, as long as the in-set (or out-set)
classification it suggests is compatible with some MIS. This is because
the frozen-node signal is sourced from the QPU's full low-energy
output, not from optimal states alone: a correct in-set node $v$
may appear in many sampled independent sets, of which the MISs containing
$v$ are only a subset. Its marginal inclusion probability is therefore
strictly larger than the probability of sampling an MIS, so the signal
qReduMIS needs survives in regimes where a bare quantum solver would
fail.

\textbf{Beyond reducible graphs.} Finally, we note that qReduMIS is not
restricted to graphs that are directly amenable to reduction. Even for
instances immune to classical reduction (e.g., 3-regular graphs, where
no vertex is isolated), removing a correctly identified frozen node
together with its neighborhood breaks the local structure and
generically yields a kernel that \emph{is} reducible, re-enabling
classical reduction on the updated kernel.

\section{Numerical experiments}
\label{sec:numerics}

We now assess the performance of qReduMIS experimentally. 
For the sake of a well-controlled testbed, we focus on hardware-native instances, foregoing the requirement for excessive compilation. 
In particular, we consider (non-planar) random union-jack (UJ) instances, as previously studied in, e.g., Refs.~\citep{ebadi:22, andrist:23, finzgar:23, kim:23}. 
We compare the performance of QAA vs a semi-greedy implementation of qReduMIS with in-set selection strategy and $20$ classical shots per instance, for the same optimized schedule \citep{perseguers:24}, and $n_{\mathrm{shots}}=1000$ throughout.
We provide results based on experiments run on the QuEra Aquila QPU available on Amazon Braket \citep{wurtz:23}. We apply post-processing logic that repairs violations of the independence constraint \citep{ebadi:22, wurtz:23}. We also perform post-selection and only consider those shots for which the lattice has been correctly filled with atoms, as done in Ref.~\citep{perseguers:24}. 

\begin{figure}
  \includegraphics[width=1.0 \columnwidth]
  {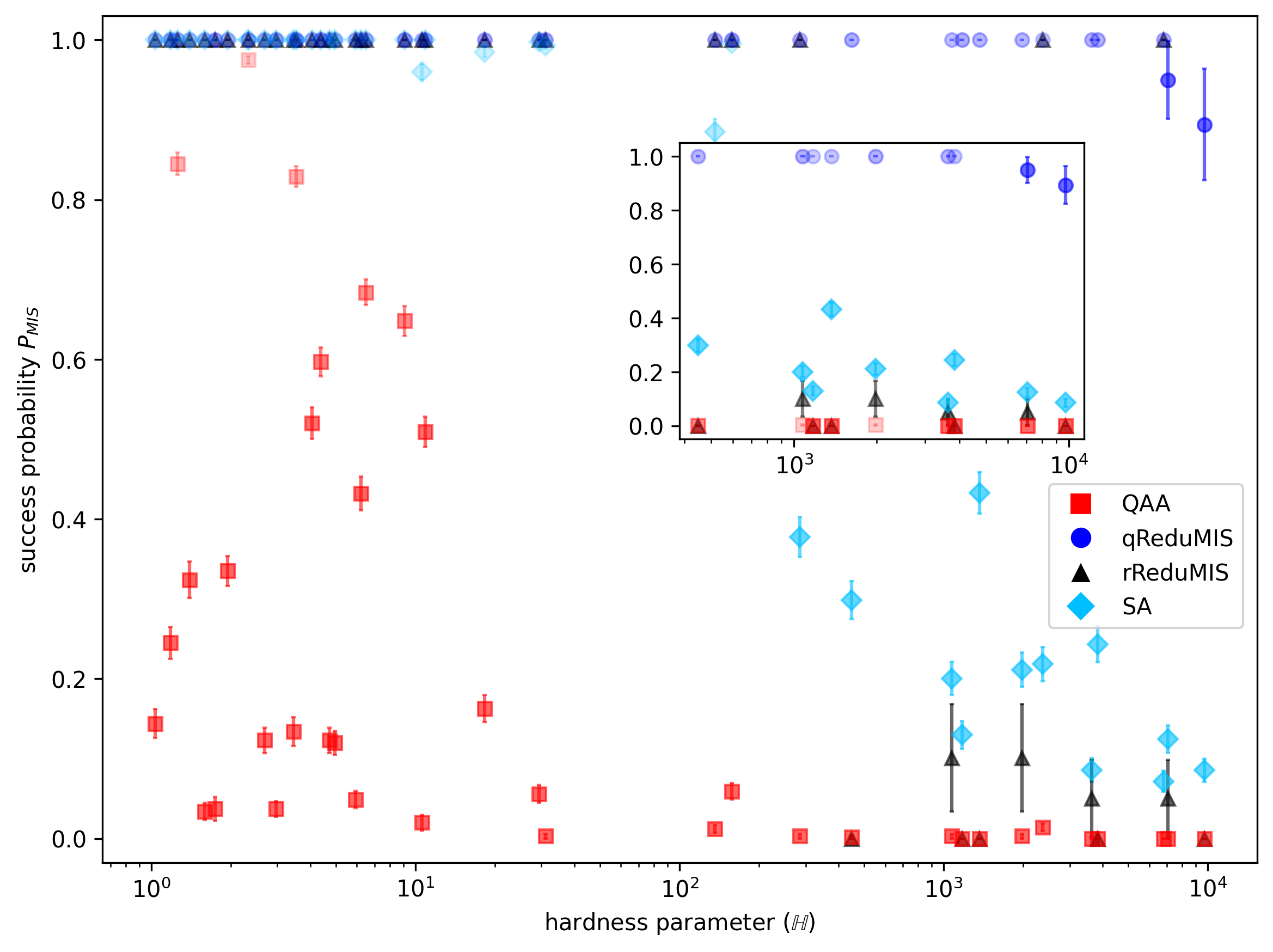}
  \caption{
  Success probability $P_{\mathrm{MIS}}$ as a function of the hardness parameter $\HP$, for QAA (red squares), classical SA (bright blue diamonds), qReduMIS (blue circles), and a random-informed baseline (black triangles). 
  Error bars refer to $90\%$ confidence intervals as extracted via bootstrapping; 
  for SA error bars are estimated via repeated Bernoulli experiments. The inset provides selected results for those nine instances (with a non-zero kernel) that are not solved by the first iteration of $\textsc{ClassicalReduce}$ alone.}
      \label{fig:performance-hp}
\end{figure}

\textbf{Example instances.} We first focus on four example instances with $n=137$ nodes and hardness ranging from $\HP \sim 1.478$ to $\HP \sim 1435$, as previously studied in Refs.~\citep{finzgar:23} and \cite{perseguers:24}. 
In line with the expected strong performance dependence on $\HP$, $P_{\mathrm{MIS}} \approx 0.478$, $P_{\mathrm{MIS}} \approx 0.004$, $P_{\mathrm{MIS}} \approx 0.019$, and $P_{\mathrm{MIS}}=0$ has been reported for those four instances, using the optimized QAA developed in Ref.~\citep{perseguers:24}. 
Our results are given in Tab.~\ref{tab:4hp}. 
To illustrate the anatomy of qReduMIS, in Fig.~\ref{fig:snapshots} we provide snapshots of the algorithm for one particular instance, showing how the algorithm can succeed in finding an MIS through repeated rounds of kernelization guided by information extracted from quantum measurements. 
Overall, we find that qReduMIS outperforms QAA on all four instances, reaching $P_{\mathrm{MIS}}=1$ on all four instances, including the hardest instance for which QAA fails to find the MIS \citep{perseguers:24}, i.e., $P_{\mathrm{MIS}}=0$. 

\textbf{Systematic experiments.} Next, we perform systematic experiments on a larger set of $39$ random UJ instances, with hardness parameters ranging over four orders of magnitude from $\HP \sim 1.03$ up to $\HP \sim 9717$, based on graphs with $n=7$ -- $231$ nodes. 
To validate the usefulness of the quantum information in guiding the kernelization process, we also consider a simple \textit{random-informed} baseline (dubbed rReduMIS), in which in-set nodes are chosen randomly from the kernel. 
As another important performance baseline, we provide results based on an optimized implementation of classical simulated annealing (SA) as described in detail in Ref.~\cite{andrist:23} and shown to be competitive with the optimized QAA results reported in Ref.~\cite{ebadi:22}.

Our results are displayed in Fig.~\ref{fig:performance-hp}. 
As expected, for QAA $P_{\mathrm{MIS}}$ is strongly suppressed for hard instances, dropping to small $P_{\mathrm{MIS}} \gtrsim 0$ for most instances with $\HP \gtrsim 10$. 
While more performant than QAA, we observe a similar performance drop for SA, as expected \cite{andrist:23}.
Conversely, qReduMIS solves most instances to optimality with $P_{\mathrm{MIS}}=1$ and maintains a non-zero success rate with average $P_{\mathrm{MIS}} \gtrsim 89\%$ throughout the testbed.
Overall, these results illustrate the benefits of introducing kernelization techniques over both classical and quantum MCMC-based methods as studied in Refs.~\cite{ebadi:22, andrist:23, perseguers:24}.
We also see that qReduMIS outperforms the random baseline rReduMIS, showing that the information extracted from the quantum backend does carry useful information. 
While our current focus has been on improvements over fundamental limitations faced by a broad set of MCMC solvers, this observation lays the groundwork for future studies assessing in detail the information obtained from the quantum sampler.
While $P_{\mathrm{MIS}} = 1.0$ for all instances with $\HP \lesssim 10^3$, we find $P_{\mathrm{MIS}} \sim 0.89$ -- $0.95$ (on average) for two hard instances with $\HP \gtrsim 10^3$, which we attribute to sub-optimal information sourced from the quantum sampler for relatively large (and, thus, harder) kernels; see Appendix~\ref{additional-numerics} for further details.
Ultimately, those performance drops may be addressed with extended reduction techniques \citep{grossmann:24} and (kernel) samplers that are insensitive to the hardness parameter, as discussed in Ref.~\citep{andrist:23}, which we leave for future research. 
Still, given that the performance of qReduMIS is imperfect ($P_{\mathrm{MIS}}<1$) only for those instances where QAA fails ($P_{\mathrm{MIS}}=0$), for our testbed we find an optimal area-under-the-curve performance boost for $P_{\mathrm{MIS}}(\mathrm{qReduMIS})/P_{\mathrm{MIS}}(\mathrm{QAA})$.

\textbf{ReduMIS powered by SA.} The broader ReduMIS framework is
agnostic to the choice of optimization subroutine used to solve the
kernel: while the original ReduMIS implementation \cite{lamm:17} is
powered by a heuristic evolutionary algorithm, in the spirit of
qReduMIS one can instead use simulated annealing (SA) as the inner
solver, which we denote \emph{SaReduMIS}. As yet another baseline, we
applied SaReduMIS across our testbed of $39$ random union-jack
instances and observed a success probability of exactly
$P_{\mathrm{MIS}} = 1$ on all instances, including the nine instances
that are not fully reducible by exact reduction rules alone. This is in
line with our expectations: in earlier work on the computational
hardness of MIS \cite{andrist:23} we showed that SA performs on par
with---or, given sufficient depth, better than---quantum annealing,
with the speedup reported in Ebadi \textit{et al.}~\cite{ebadi:22}
holding only against a restricted, simple form of SA. We therefore
expect SA to supply the classical reduction wrapper with information of
comparable quality to our quantum implementation, and our experiments
confirm this. The scientifically pertinent question our work addresses
is thus not whether classical kernels are tractable at current system
sizes---they are---but whether the reduction framework can help overcome the
fundamental scaling barrier faced by standalone quantum optimization
(specifically QAA), and our results demonstrate that it does.

\section{Conclusion and Outlook}
\label{conclusion}

In summary we have proposed and analyzed a family of quantum-informed reduction algorithms for the MIS/MWIS problems (dubbed qReduMIS) in which exact kernelization is applied in tandem with a quantum co-processor that helps guide the repeated reduction process through the identification of frozen vertices. 
The qReduMIS algorithm is hardware-agnostic and embedded in a three-tier framework with distinct application, algorithm, and hardware layers. 
Our experiments with up to $231$ qubits show that qReduMIS can help overcome fundamental performance limitations faced by (neutral atom) quantum processors. 
In future work it will be interesting to integrate backtracking techniques \citep{finzgar:23b} and expand the classical reduction toolkit \citep{grossmann:24}. 
Furthermore, one could integrate qubit compression \citep{sciorilli:25}, lightcone \citep{dupont:25}, or decomposition techniques \citep{acharya:24}, with the goal to study larger problem sizes, 
and run larger benchmarking experiments (e.g., comparing qReduMIS against classical ReduMIS) using a variety of hardware platforms such as superconducting qubits or trapped ions. \\

{\bf Code availability statement}: An open source demo version of our code is publicly available at \href{https://github.com/jpmorganchase/qredumis}{https://github.com/jpmorganchase/qredumis}.

\begin{acknowledgments}
We thank Victor Bocking, Yaroslav Kharkov, Peter Komar, Lou Romano, Peter Sceusa, Alexander Buts, Jacob Albus, Pragna Subrahmanya, and Rajagopal Ganesan for their support. H.G.K.~would like to thank Charlie Gatograbber for support during the final stages of writing this manuscript. 
\end{acknowledgments}

\section*{Disclaimer}

This paper was prepared for informational purposes with contributions from the Global Technology Applied Research center of JPMorgan Chase \& Co. This paper is not a product of the Research Department of JPMorgan Chase \& Co. or its affiliates. Neither JPMorgan Chase \& Co. nor any of its affiliates makes any explicit or implied representation or warranty and none of them accept any liability in connection with this paper, including, without limitation, with respect to the completeness, accuracy, or reliability of the information contained herein and the potential legal, compliance, tax, or accounting effects thereof. This document is not intended as investment research or investment advice, or as a recommendation, offer, or solicitation for the purchase or sale of any security, financial instrument, financial product or service, or to be used in any way for evaluating the merits of participating in any transaction.

\bibliography{bibliography}


\newpage
\appendix

\section{Additional Details}
\label{additional-information}

Here we provide further details for the qReduMIS algorithm. 
We first discuss variants of qReduMIS, such as a \textit{quantum-frugal} version to speed-up the algorithmic run time. 
Thereafter, we show how to implement qReduMIS across different quantum hardware platforms, including superconducting qubits and trapped ions. 

\begin{figure*}
  \includegraphics[width=1.0 \columnwidth]{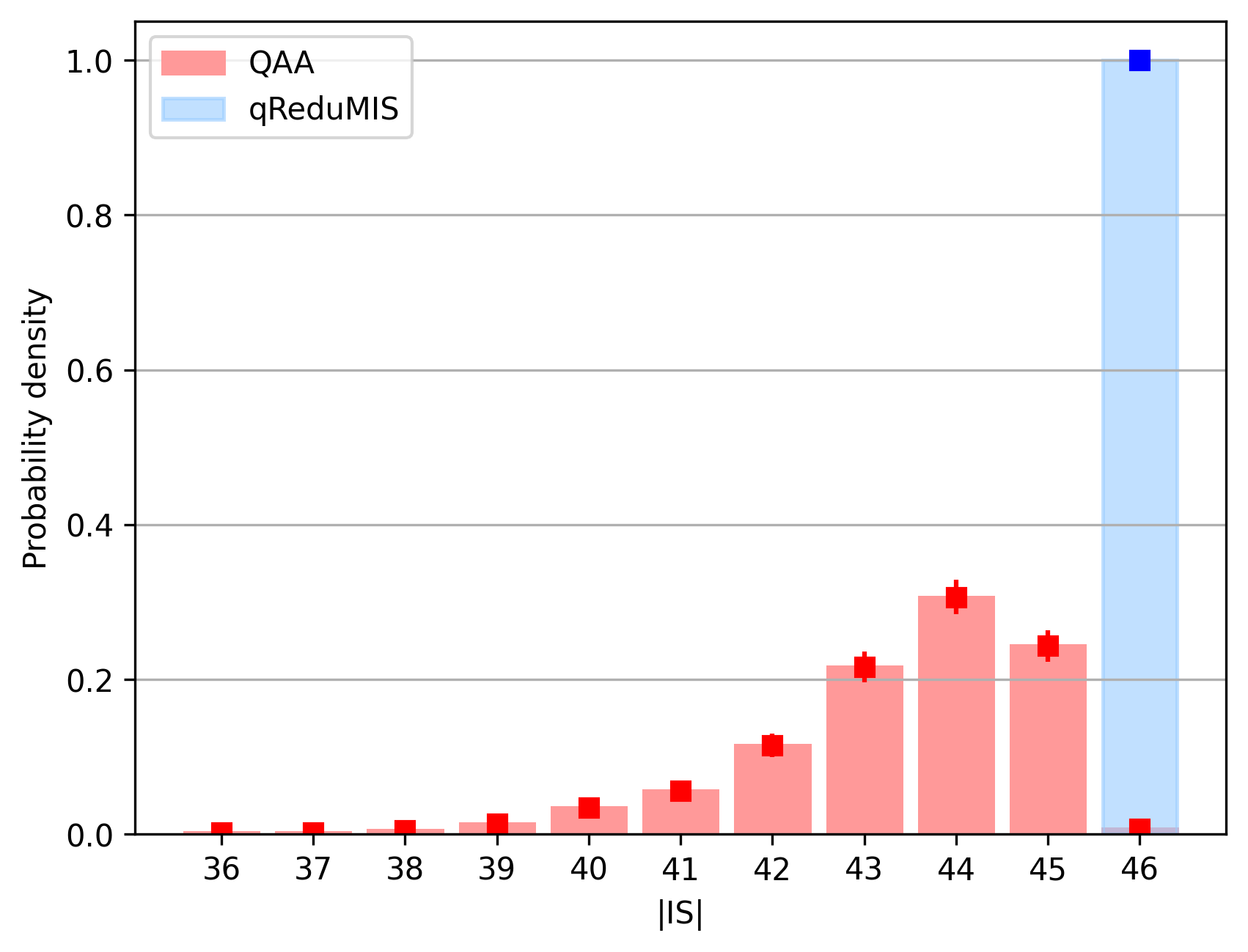}
  \includegraphics[width=1.0 \columnwidth]{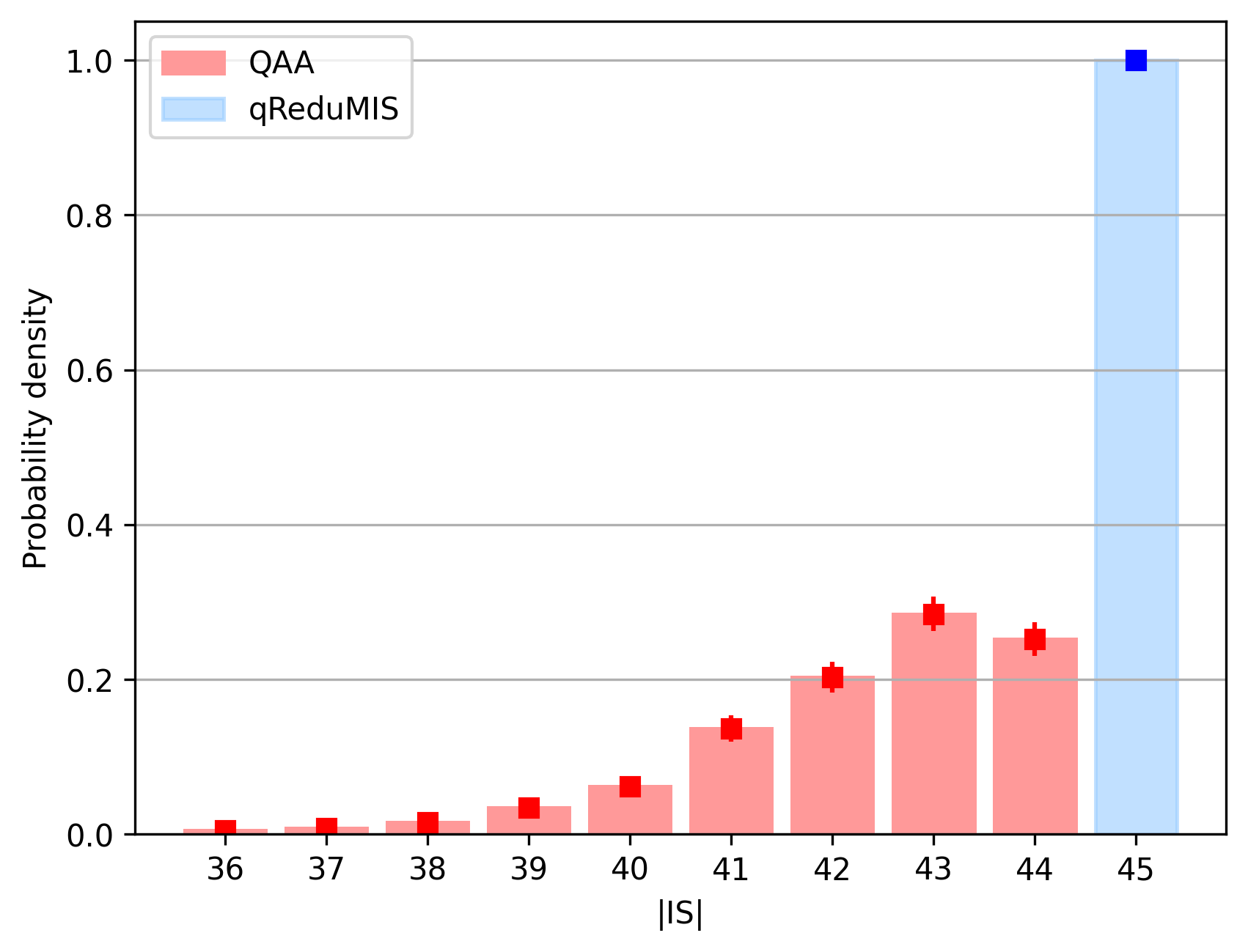}
  \caption{
  Performance as measured in terms of the independent set size $|\mathrm{IS}|$ for both QAA and qReduMIS (with in-set selection) for the two hardest instances considered in Tab.~\ref{tab:4hp} with $\HP \sim 125.5$ (left) and $\HP \sim 1435$ (right), respectively.
  The optimal MIS sizes are given in the rightmost bin.
  \textbf{(Left)} QAA finds the optimum $|\mathrm{MIS}|=46$ with a small success rate of $P_{\mathrm{MIS}} \sim 0.7\%$, displaying a broad shoulder towards suboptimal solutions, while qReduMIS finds the optimum with a success rate of $P_{\mathrm{MIS}}=100\%$. 
  \textbf{(Right)} The best solution found by QAA is $|\mathrm{IS}|=44$, while qReduMIS finds the MIS of size $|\mathrm{MIS}|=45$ with a success rate of $P_{\mathrm{MIS}} = 100\%$.
  Error bars refer to $90\%$ confidence intervals as extracted via bootstrapping, and $n_{\mathrm{shots}}=1000$ for all QPU calls.   
      \label{fig:histogram-performance}
  }
\end{figure*}

\subsection{Algorithmic Details for qReduMIS}

\textbf{Anatomy of qReduMIS.} The qReduMIS algorithm is sketched in Algorithm \ref{algo:qReduMIS}. Here, we provide some more details for its implementation. 
In each iteration, qReduMIS updates the set of selected nodes $\mathcal{S}$ that tracks nodes selected per classical reduction as well as nodes selected as (frozen) in-set kernel nodes. 
In parallel, the incumbent $\mathcal{W}$ is also updated in every iteration whenever a better solution has been found, allowing us to track (potentially high-quality) global solution candidates already in the first iteration, even when the kernel $\mathcal{K}$ may still be relatively large and $|\mathcal{S}|$ relatively small.  
As such, the QPU's output in the form of candidate sets $\{\mathcal{I}_{n}\}$ is used in two ways, to establish a global incumbent solution and to identify frozen nodes as to unblock the kernelization process. 
In this way qReduMIS proceeds towards large independent sets in two complementary ways, with the incumbent $\mathcal{W}$ securing a global solution candidate even when the algorithm is stopped early. 

\textbf{Variants of qReduMIS.} For $\lambda=K_{\mathrm{RCL}}=1$ the semi-greedy implementation of qReduMIS discussed in the main text simplifies to a simple greedy variant of qReduMIS. 
Generalizations of our semi-greedy implementation (with uniform random sampling from top-$K$ candidates) to alternative sampling strategies, such as top-$p$ (nucleus) sampling \citep{holtzman:20}, should be straightforward. 
In semi-greedy qReduMIS with $R$ parallel runs (repetitions) the total number of QPU calls amounts to $N_{\mathrm{QPU}} \sim R \cdot D$.   
However, this number can be reduced to $N_{\mathrm{QPU}} \sim D$ in a \textit{quantum-frugal} variant of qReduMIS, whereby the inexpensive classical reduction is run for all candidate nodes in the restricted candidate list RCL (as to validate multiple scenarios based on quantum information) before one greedily uses the option with the largest reduction footprint.

\textbf{qReduMIS run time.} The qReduMIS run time may be further reduced by keeping the number of shots $n_{\mathrm{shots}}$ low, because high-resolution measurements are not necessarily required, akin to the use of stochastic gradient descent over gradient descent. 
Moreover, the run time depends heavily on the quantum device used and its characteristic cycle time $\tau$. 
For Rydberg atom arrays, we currently have relatively slow cycle times with $\tau \sim 0.1\mathrm{s}$, amounting to $\sim 10$ samples per second. 
Conversely, with superconducting devices one can achieve more than $6\times 10^{4}$ circuit repetitions per second \citep{neven:24}, resulting in much faster cycle times with $\tau \sim 1.6 \times 10^{-5} \mathrm{s}$.  

\textbf{Robustness.} By design, qReduMIS is robust against errors in that it can succeed in finding the MIS for the original input $\mathcal{G}$ even if the QPU fails to find the MIS for the kernels $\mathcal{K}$, as long as the in-set (or out-set) classification suggested by the QPU is compatible with some MIS solution. 
For typical, sufficiently dense problems of interest, the majority of vertices will be out-set nodes resulting in an imbalanced classification problem in which the correct prediction of some out-set nodes is expected to be easier than the correct prediction of in-set nodes. 
For example, for the instances studied in Tab.~\ref{tab:4hp} with $n=137$ and typical MIS size of $\sim 46$ the average in-set probability is approximately $\bar{p}_{\mathrm{in}} \sim 1/3$, while the average out-set probability is approximately $\bar{p}_{\mathrm{out}} \sim 2/3$.
However, removal of in-set nodes comes with a larger reduction footprint and thus typically requires a smaller number of qReduMIS iterations. 
Empirically, we observe that the two effects roughly balance each other out, leading to approximately similar performance for in-set and out-set selection strategy. 
We leave a more detailed analysis thereof, along with the development and analysis of alternative selection strategies, for future research. 

\textbf{Loading errors.} In particular, qReduMIS can also improve the robustness against atom loading errors for Rydberg devices. 
Specifically, for the QuEra Aquila device, the typical probability of failing to occupy a given (single) site is $\epsilon \sim 0.007$ \citep{wurtz:23}. 
The probability to correctly load a desired atom arrangement with $n$ atoms is then $p_{\mathrm{load}} \approx (1-\epsilon)^n$. 
We have addressed this source of error through proper post-selection, noting that the fraction of valid samples can be much larger for qReduMIS over QAA, given that the kernel graph is typically much smaller than the original graph.
For example, for the hardest instance in Tab.~\ref{tab:4hp}, we have $p_{\mathrm{load}} \sim 38\%$ for QAA with $n=137$, while $p_{\mathrm{load}} \sim 77\%$ for correctly loading the (first) kernel with $n=37$ nodes within qReduMIS.

\subsection{Hardware Layer}
\label{app:hardware}

The qReduMIS framework is hardware-agnostic. 
In the main text we have focused on implementations based on Rydberg atom arrays. 
Here, we now show how qReduMIS could be backed by digital (gate-based) devices run, for example, on superconducting (SC) qubits or trapped ions, or alternative SC-based quantum annealers. In order to use these backends within the qReduMIS framework, one needs to specify the unitary $\mathcal{U}(\theta)$ that implements the desired dynamics in Eq.~(\ref{eq:quantum-state}). Below, we describe this process for both the quantum approximate optimization algorithm using digital architectures, and quantum annealing using superconducting qubits.

\textbf{QAOA.} The quantum approximate optimization algorithm (QAOA) is one of the most widely studied quantum algorithms for solving combinatorial optimization problems on \textit{gate-based}, near-term quantum devices \citep{farhi:14}, with first experimental demonstrations run on both superconducting qubits \citep{otterbach:17, harrigan:21} and trapped ions \citep{pagano:20}.   
In the case of QAOA, the unitary $\mathcal{U}(\theta)$ in Eq.~(\ref{eq:quantum-state}) is given by $\mathcal{U}(\theta)=\Pi_{l=1}^{p}U_{\mathrm{mix}}(\beta_{l}) U_{\mathrm{cost}}(\gamma_{l})$, involving a series of $p$ layers alternating cost and mixing unitaries, $U_{\mathrm{cost}}(\gamma_{l})=\exp(-i\gamma_{l}\hat{H}_{\mathrm{cost}})$ and $U_{\mathrm{mix}}(\beta_{l})=\exp(-i\beta_{l}\hat{H}_{\mathrm{mix}})$, respectively, which are generated by $\hat{H}_{\mathrm{cost}}$ and $\hat{H}_{\mathrm{mix}}=\sum_{i}\hat{X}_{i}$, respectively. The parameters $\theta=(\boldsymbol{\gamma}, \boldsymbol{\beta})$ can either be optimized in an outer (closed) loop, inferred from parameter setting heuristics \citep{zhou:20, he:24, wurtz:21}, or pre-optimized \citep{augustino:24}.
Finally, note that one could also adopt generalizations of QAOA such as recursive QAOA (RQAOA) \citep{bravyi:20}.

\textbf{Superconducting annealers.} Alternatively, when adopting superconducting (SC) quantum annealers, e.g., as provided by D-Wave Systems Inc.~\citep{johnson:11, kadowaki:98, das:08, hauke:20}, the unitary $\mathcal{U}(\theta)$ in Eq.~(\ref{eq:quantum-state}) is generated by the time-dependent Hamiltonian 
$\hat{H}(t)=(1-\theta)\hat{H}_{\mathrm{mix}}+\theta \hat{H}_{\mathrm{cost}}$, where $\theta=\theta(t)$ defines the annealing schedule with $\theta(0)=0$ and $\theta(\tau_{f})=1$ for a total anneal time $\tau_{f} \sim 10\mu\mathrm{s}$.
More general protocols can involve techniques such as counterdiabatic driving \citep{campo:13}, reverse annealing \citep{ohkuwa:18}, Bayesian optimization \citep{finzgar:23}, or catalyst terms \citep{ghosh:24, nutricati:24}.

\section{Additional Numerical Results}
\label{additional-numerics}

In this Appendix, we provide further details for our hardware experiments and additional numerical results, complementing the results shown in the main text.

\begin{table}
\caption{
Success probabilities $P_{\mathrm{MIS}}$ (including error estimates) achieved with QAA and qReduMIS (for both an in-set and out-set selection strategy) using the QuEra Aquila QPU \citep{wurtz:23} for four test instances (all with $n=137$ vertices) referenced by their hardness parameters $\HP$.
\label{tab:4hp-extended}} 
 \begin{tabular}{ l c c c c } 
 \hline
 \hline
 Algorithm \hfill$\HP$ & $\sim1.478$ & $\sim 14.12$ & $\sim 125.5$ & $\sim 1435$ \\ [0.5ex] 
 \hline
 QAA & $0.35(2)$ & $0.008(5)$ & $0.007(4)$ & 0.0   \\ 
 qReduMIS (out-set)& 1.0 & 1.0 & 1.0 & 1.0\\ 
 qReduMIS (in-set)& 1.0 & 1.0 & 1.0 & 1.0  \\ [1ex]
 \hline
 \hline
 \end{tabular}
\end{table}

\textbf{Algorithm parameters}. In our experiments we have set $\lambda=1$, and $K_{\mathrm{RCL}} = 0.4 |\mathcal{K}|$, where $|\mathcal{K}|$ refers to the number of kernel nodes. 
Measurements have been filtered to solution candidates $\{\mathcal{I}_{n}\}$ with the largest and second largest independent set size found.

\textbf{Comparison to simulated annealing}. Our implementation of SA is the one developed in Ref.~\cite{andrist:23}. Note that our implementation differs from the implementation of SA presented in \cite{ebadi:22} in the following ways: (i) our implementation involves only moves compatible with the (hard) independence criterion such that only the feasible space is
searched and is not based on a (soft) penalty model, (ii) proposal probabilities are biased towards additions and exchanges to increase acceptance, and (iii) we use a geometric cooling schedule in combination with frequent restarts, as opposed to a constant low temperature as used in Ref.~\cite{ebadi:22}. These changes allow the implementation of SA in \cite{andrist:23} to achieve higher success probability compared to Ref.~\cite{ebadi:22}. 
We estimate the $90\%$ confidence interval for our results using the statistical variance from repeated Bernoulli trials, where each trial is considered a success if the optimal solution is found and a failure otherwise. The variance for $N_{B}$ Bernoulli trials is given by $\hat{p}(1-\hat{p})/N_{B}$, where $\hat{p}$ represents the observed probability of success. To estimate the $90\%$ confidence interval, we multiply the standard error by 1.645, which is the z-score corresponding to the central $90\%$ of the standard normal distribution.

\textbf{Comparison to SaReduMIS.} We also implemented the ReduMIS framework powered with classical SA as kernel solver, 
replacing the QAA sampler from qReduMIS with SA. Note that the original ReduMIS \cite{lamm:17} algorithm comes with a larger set 
of reduction rules but for this comparison we employed the same classical reduction rule based on clique-based removals. 
To make the comparison fair, we employed 500 replicas and two steps in the SA implementation such that the total computational 
budget was 1000, on par with the QAA component of qReduMIS.

\textbf{Hardware parameters.} For our hardware experiments we adopt the optimized schedules developed in Ref.~\citep{perseguers:24}, with a maximum Rabi frequency $\Omega_{\mathrm{max}}/2\pi = 2.5\mathrm{MHz}$ and total annealing time $\tau_{f}=4\mu \mathrm{s}$.
The lattice spacing $a$ was set to $a=5.45 \mu \mathrm{m}$ for most of the problem instances, ensuring that the Rydberg blockade radius $R_b \equiv (C_{6}/\Omega_{\mathrm{max}})^{1/6}$ is compatible with the desired UJ connectivity, i.e., $\sqrt{2} \leq R_{b}/{a} < 2$ for $R_{b}/{a}\sim 1.5$.
Given device geometry constraints in the form of limited maximum site pattern width and height \citep{wurtz:23}, for large instances we set $a=(4.45 - 5.0) \mu \mathrm{m}$ when necessary, amounting to $R_{b}/{a}\sim 1.7 - 1.9$, well within the desired UJ regime.

\begin{figure}
  \includegraphics[width=1.0 \columnwidth]{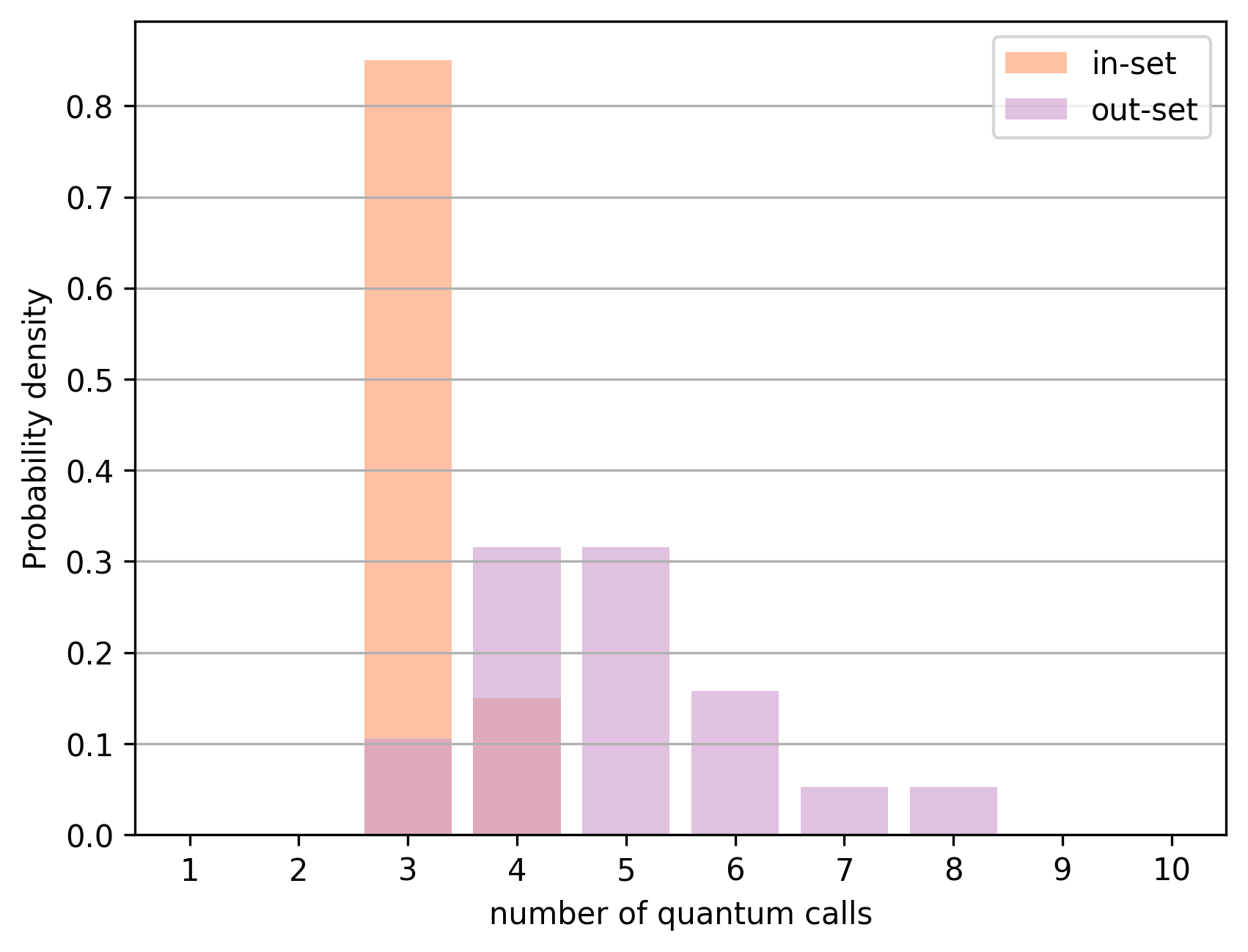}
  \includegraphics[width=1.0 \columnwidth]{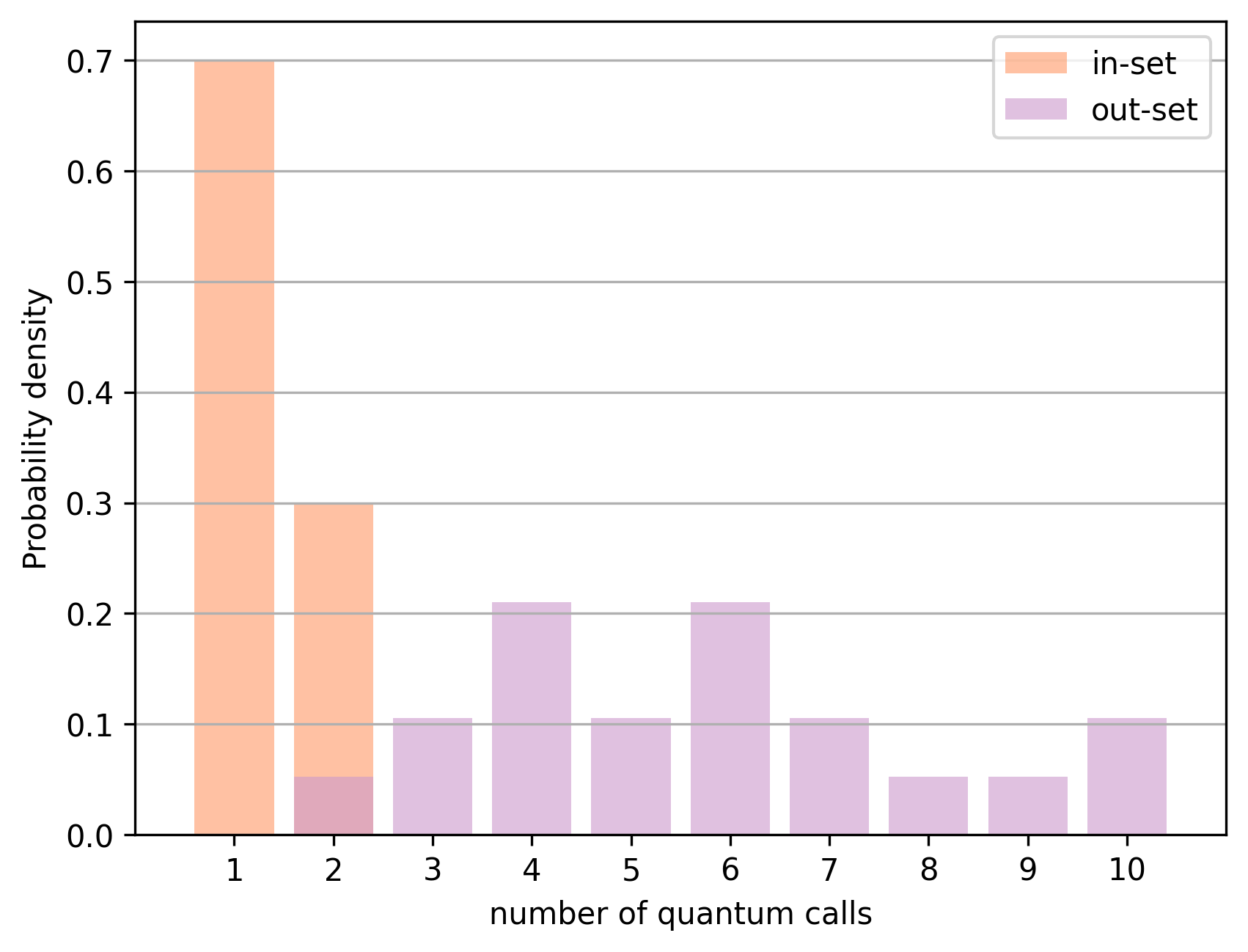}
  \caption{
  Number of QPU calls for both in-set and out-set selection strategy (with $\lambda=1$) for the two hardest instances considered in Tab.~\ref{tab:4hp} with $\HP \sim 125.5$ (top) and $\HP \sim 1435$ (bottom), respectively. 
      \label{fig:histogram-qpu-calls}
  }
\end{figure}

\begin{figure}
  \includegraphics[width=1.0 \columnwidth]{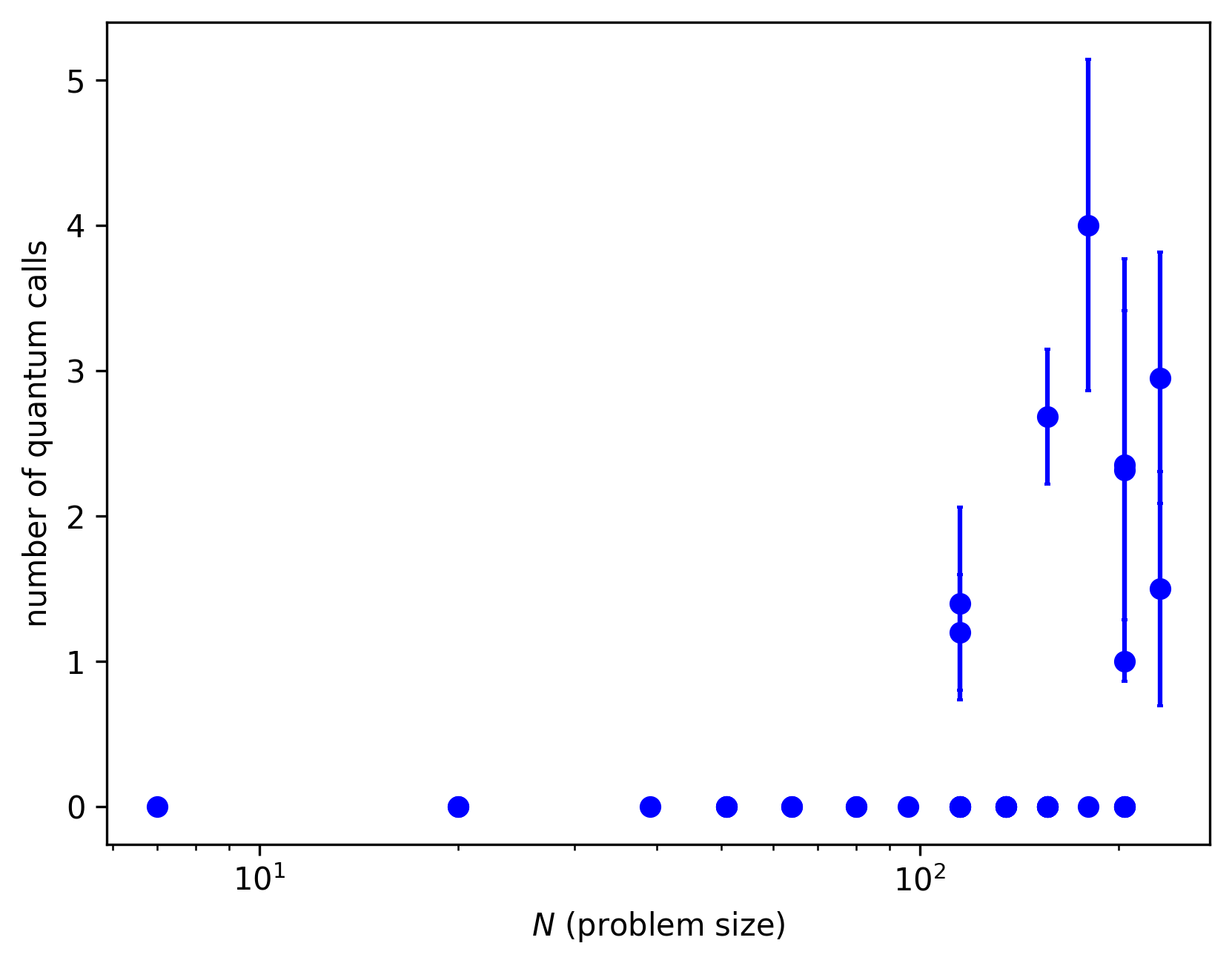}
  \caption{
  Number of QPU calls for qReduMIS with in-set selection across the larger testbed with $39$ instances with up to $231$ nodes. 
  Note that the number of QPU calls is zero for the majority of instances, thus providing a QPU resource efficiency improvement over vanilla QAA for this set of (fully reducible) instances.
      \label{fig:qpu-calls}
  }
\end{figure}

\textbf{Example instances.} We first provide an extended version of Tab.~\ref{tab:4hp} that features error bars (as extracted from bootstrapping) and provides results for both in-set and out-set selection strategies; see Tab.~\ref{tab:4hp-extended}. 
We observe comparable performance for the two selection strategies. 
In addition, complete histograms featuring all shots (QAA) and repetitions (qReduMIS) for the two hardest instances considered in Tab.~\ref{tab:4hp} are displayed in Fig.~\ref{fig:histogram-performance}. 
We see that with qReduMIS the probability is shifted to larger independent set sizes, indicating that performance is improved not only in the best case, but also on average.  

\textbf{Number of QPU calls.} In Fig.~\ref{fig:histogram-qpu-calls}, we provide histograms showing the number of QPU calls (across $R=20$ runs) for both in-set and out-set selection strategies for the two hardest instances considered in Tab.~\ref{tab:4hp}. 
We find that a small number of calls ($\lesssim 10$) is sufficient for these instances with $n=137$ nodes, in line with the expected (moderate) logarithmic increase of algorithmic depth $D$ with system size $n$.  
Moreover, we find that in-set selection typically requires a smaller number of QPU calls, likely due to its larger removal footprint, because we only remove one (frozen) node at a time (for $\lambda=1$) within the out-set strategy, but we remove one frozen node together with its kernel neighborhood when adopting the in-set strategy. 
As such, qReduMIS is expected to reach an empty kernel faster when adopting the in-set strategy.
Finally, we note that the number of QPU calls can be reduced with a simple extension of our selection strategy in which we select one node per kernel component which allows to (potentially) unblock all kernel components in parallel.

Likewise, the number of QPU calls for qReduMIS with in-set selection across the larger testbed underlying Fig.~\ref{fig:performance-hp} is shown in Fig.~\ref{fig:qpu-calls}. 
We find that small UJ instances with $n \lesssim 100$ nodes tend to be fully reducible, and thus do not require any QPU calls \citep{schuetz:24}, 
thus providing a QPU resource efficiency improvement over vanilla QAA for this set of (fully reducible) instances. 
Some of the larger instances with $n \gtrsim 100$ nodes may have a non-empty kernel, and thus may require calls to the QPU. However, compared to vanilla QAA, the overhead is small in that a small number of calls is sufficient to reach termination of qReduMIS. 

We also observe a relatively strong Pearson correlation coefficient $\sim -0.92$ between the number of QPU calls and the size of the first reduction, i.e., the larger the first classical reduction, the smaller the average number of QPU calls. 
The number of QPU calls is anti-correlated with the performance measure $P_{\mathrm{MIS}}$, as confirmed by a relatively strong Pearson correlation coefficient of $\sim -0.40$. 
Overall, the emerging rationale can be simply stated as follows: 
The larger the problem size $n$, the smaller the average reduction, the larger the kernel, the larger the number of QPU calls, the smaller the average $P_{\mathrm{MIS}}$.

\textbf{Fixed shot budget.} As explained in the main text, our main experiments used the same number of quantum shots in QAA as in each execution of the quantum backend within the full qReduMIS solve. In some cases, qReduMIS requires more than one quantum call (i.e., execution of the quantum backend), as shown in Fig. \ref{fig:qpu-calls}, such that qReduMIS could incur a larger total number of quantum shots than QAA. To further compare qReduMIS with QAA, we also ran additional experiments with a fixed \textit{total} number of quantum shots. For example, we observed that instances with $\HP\gtrsim 10^{2}$ have a larger number of quantum calls. Note, however, that the number of qubits (nodes) required for the kernels solved by the quantum annealing system in qReduMIS is smaller than that of standalone QAA for the full graph, as we apply classical reductions effectively reducing the size of the problem, thus reducing the QPU resource requirements in terms of the number of qubits.

To study the trade-off between the number of quantum calls and the success probability, we performed additional experiments comparing QAA and qReduMIS (via the noiseless tensor network simulator) for different \textit{total} numbers of quantum shots. The total number quantum shots for standalone QAA is simply the number of shots that the program ran, whereas for qReduMIS the total number of quantum shots is the sum over all the quantum shots used across all calls to the QPU.

We performed these additional experiments for the two hardest instances considered in Tab.~\ref{tab:4hp}, with $\HP \sim 125.5$ and $\HP \sim 1435$. For the first instance, we set the number of quantum shots per iteration of qReduMIS to $125$ and $250$, respectively, and we calculate the success probabilities across $20$ classical shots, and we report the largest value of total quantum shots (i.e, the worst case). In all cases, the success probability obtained by qReduMIS is $1$. In contrast, QAA achieves a success probability of $0$, even when running for $1000$ total quantum shots. We report these results in Table \ref{tab:quantum_shots}. We attribute the lack of decay in success probability for qReduMIS when reducing the number of quantum shots per quantum call to the fact that the bulk of the heavy lifting in the algorithm is classical reduction, and the QPU is only leveraged to identify frozen nodes that unblock subsequent classical reduction. Performing this task requires less precision than identifying the actual problem solution, as one does in standard QAA. Note that the average number of QPU calls does not change much when varying the number of quantum shots (i.e., $3.1$ calls for 500 total of quantum shots, and $3.05$ calls for $1000$ quantum shots).

For the hardest instance, we ran qReduMIS with $250$ and $500$ quantum shots per iteration, incurring a total of $500$ and $1000$ quantum shots (in the worse case). In Table \ref{tab:quantum_shots} we see that in this case there is an improvement in the success probability as we increase the number of quantum shots per iteration. 
We observe that $P_\text{MIS}^\text{worst}$ (the success probability when we restrict to only the trials (i.e., classical shots) that incurred the worst case number of total quantum shots) is significantly improved when increasing the quantum shots per iteration, as expected. When using $500$ quantum shots per QPU call, out of the 20 classical shots, 12 performed three classical reductions (two QPU calls). This number is very similar when using $250$ quantum shots per QPU call, with $11$ (out of the 20 classical shots) shots performing three classical reductions.

Overall, based on our experiments we conclude that qReduMIS can outperform QAA even when the budget of total quantum shots is fixed and held constant.

\begin{table}
\caption{
Success probabilities $P_{\mathrm{MIS}}$ achieved for QAA and qReduMIS (using the in-set selection strategy) for different total numbers of quantum shots ($qshots$) when solving two of the hardest test instances (with $n=137$ vertices), labeled by hardness parameter $\HP$. Note that in QAA the total quantum shots is simply the number of shots used in the program execution, while in qReduMIS the total is given by the sum of all quantum shots the QPU executed in all iterations of the algorithm. Here, $P_\text{MIS}^\text{worst}$ is the success probability for qReduMIS when restricted to only those trials (i.e., classical shots) for which the worst case number of total quantum shots was required. 
}

 \begin{tabular}{ l|  c c c | c c c } 
 \hline
 \hline
 &\multicolumn{3}{c|}{$\HP \sim 125.5$} & \multicolumn{3}{c}{$\HP \sim 1435$} \\ [0.5ex]
 \hline
 \hline
 qshots & QAA & qReduMIS & $P_\text{MIS}^\text{worst}$ & QAA & qReduMIS & $P_\text{MIS}^\text{worst}$ \\ [0.5ex] 
 \hline
 500 & 0.0 & 1.0 & 1.0 & 0.0 & 0.6 & 0.27 \\  
 1000 & 0.0 & 1.0 & 1.0 & 0.0  & 0.80 & 0.75 \\ [1ex] 
 \hline
 \hline
 \end{tabular}
\label{tab:quantum_shots}
\end{table}

\begin{table}
\caption{
Success probabilities $P_{\mathrm{MIS}}$ (including error estimates corresponding to 90$\%$ confidence intervals extracted via bootstrapping) obtained with QAA and qReduMIS (for in-set selection strategy) for different number of number of frozen nodes ($\lambda$) for two of the hardest test instances (all with $n=137$ vertices) referenced by their hardness parameters $\HP$. Note that the success probability corresponding to standard QAA is exactly $0$ for both instances. We report on the number quantum calls $qcalls$, which is the number of times the QPU was called to unblock reduction. Note that for the instance whose $\HP \sim 125.5$ results for $\lambda=3$ are not included because an intermediate kernel has a smaller size of MIS, as discussed in the text. }

 \begin{tabular}{ r@{\hspace{7.5pt}}|  c c | c c } 
 \hline
 \hline
 & \multicolumn{2}{c|}{$\HP \sim 125.5$} & \multicolumn{2}{c}{$\HP \sim 1435$} \\ [0.5ex]
 \hline
 \hline
  $\lambda$ & qReduMIS & qcalls & qReduMIS & qcalls \\ [0.5ex] 
 \hline
 1 & 1.0 & 3.2(4) & 0.6(1)  & 1.6(5) \\ 
 2 & 1.0 & 3.0    & 0.4(1)  & 1.2(4) \\ 
 3 & -   & -      & 0.20(9) & 1.2(4) \\[1ex]
 \hline
 \hline
 \end{tabular}
\label{tab:frozen_nodes}
\end{table}

\textbf{Number of frozen nodes.} As discussed in the main text, one of the hyper-parameters in the qReduMIS algorithm is the number of frozen nodes to be identified by the quantum backend either for inclusion in the candidate independent set (in-set) or exclusion (out-set). For the benchmarks shown in the main text, we set $\lambda=1$. However, one can select more than one node with high probability to be included or excluded from the set. In the case of in-set selection, identified nodes are included in the candidate set, and they are removed from the kernel along with their neighbors. To understand the effect of changing this hyperparameter, we analyzed the success probability of the qReduMIS algorithm while varying the number of frozen node in the case of in-set selection using the noiseless tensor network simulator for the two hardest instances considered in Tab.~\ref{tab:4hp} ($\HP \sim 125.5$ and $\HP \sim 1435$). 
Using the output from the TN simulator, we can generalize the concept of frozen nodes to larger subsets of nodes.
For example, if $\lambda=2$, we can obtain the probability of any possible pair of nodes occurring in the output. Ordering these probabilities, we can then select the set of nodes (with size $\lambda$) with the highest probability of occurring, which we return in place of a single frozen node. If multiple such sets occur with the same (largest) probability, we randomly select one to return. In the in-set strategy, these selected nodes are removed together with their neighbors. We study the performance of qReduMIS (in terms of the success probability and the number of quantum calls, i.e., the number of times the quantum backend is utilized per execution of qReduMIS) for different values of $\lambda$ across $20$ classical shots. We set the number of quantum shots per iteration of qReduMIS to $250$, as we have shown that for this value the total number of quantum shots is less than or equal to $1000$ (as used for QAA). We report the results in Table \ref{tab:frozen_nodes}. For the easier of the two instances, we observe that neither the success probability nor the number of quantum calls significantly change when increasing $\lambda$. Conversely, for the harder instance we observe a significant reduction in the success probability with increasing $\lambda$, though no significant change in the required number of quantum calls. These results suggest that transferring responsibility from the \textit{exact} classical methods to the \textit{heuristic} quantum backend can lead to reduced performance, even though the kernel is simplified more for larger $\lambda$. Overall, with larger values of $\lambda$ one can reduce the number of quantum calls (given the larger heuristic node removal), albeit at the expense of potential performance drops (given that exact reduction is traded for heuristic removal), as expected. 

In order to increase the probability of success for larger $\lambda$, one would need more quantum shots to have higher confidence that the selected nodes are indeed within the MIS. However, selecting $\lambda$ too large (e.g., larger than $|\mathrm{MIS}|$ for any given kernel) can lead to failure, as the algorithm is asked to produce $\lambda$ nodes that are likely to all be within the MIS, but only $|\mathrm{MIS}|<\lambda$ nodes are possible. We see an example of this phenomenon for the case of $\HP\sim 125.5$ in Table \ref{tab:frozen_nodes}, in which an intermediate kernel has $|\mathrm{MIS}|=2$, but $\lambda=3$. We observed (across many classical seeds) that, after the second iteration of qReduMIS, the resulting kernel has five nodes and is 3-regular with $|\text{MIS}|=2$. Thus, it was not possible to remove $\lambda=3$ nodes through in-set identification. For this reason we did not include results for $\lambda=3$ in the table. Note that, if the kernel graph is disconnected with $\lambda$ disconnected components, one can immediately identify frozen nodes in each disconnected subgraph independently without any compromise, as these frozen nodes can be correctly identified individually.

\FloatBarrier
\end{document}